\documentclass{appolb}
\usepackage{epsfig}
\usepackage{amsmath}
\usepackage[english]{babel}
\usepackage[T1]{fontenc}
\usepackage[utf8]{inputenc}
\usepackage{times}
\usepackage[numbers,sort&compress]{natbib}
\usepackage{setspace}

\newcommand{\ep}{$\eta'$}
\newcommand{\epw}{$\Gamma_{\eta'}$}

\newcommand{\cc}{\mbox{COSY--11}}
\begin{document}
\title{STUDY OF THE $\eta'$ MESON STRUCTURE, WIDTH AND INTERACTIONS WITH NUCLEONS AT COSY--11%
\thanks{Presented at Excited QCD, 31.02.-06.03.2010 in Stara Lesna, Slovakia}%
}
\author{E.~Czerwi\'nski$^{a,b,c}$, J.~Klaja$^b$, P.~Klaja$^b$, and P.~Moskal$^{a,b}$
\address{$^a$~Institute of Physics, Jagiellonian University, 30-059 Cracow, Poland}
\address{$^b$~Institute for Nuclear Physics and Juelich Center for Hadron Physics,\\Research Center J\"ulich, 52425 Juelich, Germany}
\address{$^c$~INFN, Laboratori Nazionali di Frascati, 00044 Frascati, Italy}
}
\maketitle
\begin{abstract}
We present results on the isospin dependence of the $\eta'$ production cross-section in nucleon--nucleon collisions,
as well as the results of comparative analysis of the invariant mass
distributions for the $pp\to pp\eta'$ and $pp\to pp\eta$ reactions in the context of the proton--$\eta$ and
proton--$\eta'$ interaction. Additionally, the value of the total width of the $\eta'$ is reported
as derived directly from the measurement of the mass distribution and an explanation
of the experimental technique used in order to achieve a precision about an order of magnitude better then former experiments is included.
\end{abstract}
  
\section{Experimental setup}
\begin{figure}[t]
\begin{center}
    \includegraphics[width=0.55\textwidth]{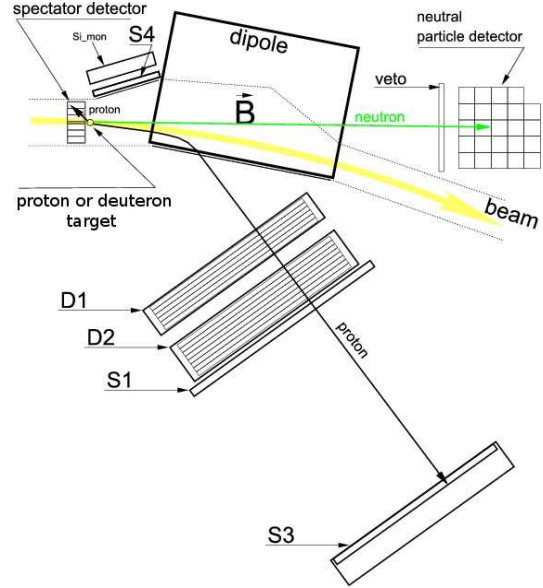}
\end{center}
 \caption{
         Schematic view of the \cc\ detector setup (top view).
         S1, S3, S4 denote scintillator detectors, D1, D2 indicate drift chambers and  Si stands
         for the silicon-pad detector. An array of silicon pad detectors (spectator detector) is used for the registration
         of the spectator protons. Neutrons are registered in the neutral particle detector.
         Detectors S4 and Si were used for the measurement
         of elastically scattered protons needed for monitoring purposes~\cite{Moskal5,ErykPhD,jklaja-phd,pk_phd}.
         Example of $pd\to pnp_{spectator}X$ reaction is presented.
         }
 \label{c11}
\end{figure}

The reported experiments have been performed in the Research Centre J{\"u}lich at the cooler synchrotron
COSY~\cite{Maier} by means of the COSY--11 detector system~\cite{Brauksiepe} presented in Fig.~\ref{c11}.
The collision of a proton from the COSY beam with a proton or deuteron cluster target may cause an $\eta'$ meson creation.
In that case all outgoing nucleons have been registered by the COSY--11 detectors, whereas
for the $\eta^{\prime}$ meson identification the missing mass technique was applied.
%joanna%%%%%%%%%%%%%%%%%%%%%%%%%%%%%%%%%%%%%%%%%%%%%%%%%%%%%%%%%%%%%%%%%%%%%%%%%%%%%%%%%%%%
\section{Production of the \ep\ meson in the $pn\to pn\eta'$ reaction}
 The main goal of this experiment was the determination of the excitation function
 for the quasi-free $pn \to pn\eta^{\prime}$ reaction
 near the kinematical threshold. The motivation was the
 comparison of the $pp \to pp\eta^{\prime}$ and
 $pn \to pn\eta^{\prime}$ total cross-sections in order to learn about the production mechanism
 of the $\eta^{\prime}$ meson in the channels of isospin 1 and 0, and to investigate
 aspects of the gluonium component of the $\eta^{\prime}$ meson.

The ratio $R_{\eta^{\prime}}~=~{{\sigma(pn \to pn\eta^{\prime})} / {\sigma(pp \to pp\eta^{\prime})}}$
has not been measured so far,
and the existing predictions differ drastically
depending on the model.
Cao and Lee~\cite{cao-prc781}
assumed,
by analogy to the production of the $\eta$ meson,
that the production of
the $\eta^{\prime}$ meson proceeds
dominantly via the S$_{11}$(1535) resonance.
As a consequence, they predicted within an effective Lagrangian approach
a $R_{\eta^{\prime}}$ value
equal to the experimentally established $R_{\eta}~=~{{\sigma(pn \to pn\eta)} / {\sigma(pp \to pp\eta)}}$ value.
In contrast,
Kaptari and K\"ampfer~\cite{kampfer-ep1}
predicted  a value of $R_{\eta^{\prime}}$
close to $\sim$1.5 in the kinematic range of the COSY--11 experiment
with the dominant contribution coming from the  meson conversion currents.
In the extreme scenario of glue-induced production saturating
the $\eta^{\prime}$ production cross-section, the ratio
$R_{\eta^{\prime}}$
would approach unity
after correcting for the final state interaction between the two outgoing nucleons.

\begin{figure}[h]
\begin{center}
\includegraphics[width=0.55\textwidth,angle=0]{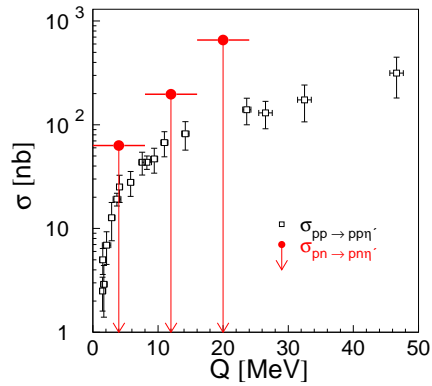}
\end{center}
\caption{Total cross-sections for the $pp \to pp\eta^{\prime}$
         reaction as a function of the excess energy (open squares).
         Upper limit for the total cross-section for the $pn \to pn\eta^{\prime}$
         reaction as a function of the excess energy (dots).
}
\label{cross_pn}
\end{figure}

Detailed description of measurement and data analysis
is presented in \cite{jklaja-phd,jklaja_prc}.
The result is shown in Fig.~\ref{cross_pn}.  The horizontal bars represents the intervals of the excess
energy, for which the upper limit of the total cross-section was calculated.
The total cross-section for the $pp \to pp\eta^{\prime}$ reaction was measured in
previous experiments~\cite{moskal-plb474,khoukaz-epj,balestra-plb491,moskal-prl80,hibou-plb438}. It reveals a strong excess energy dependence,
especially very close to threshold. This dependence must be taken into account when comparing
to the results for the $pn \to pn\eta^{\prime}$ reaction which were established for 8~MeV
excess energy intervals. Therefore, for a given interval of excess energy, we have determined
the mean value of the total cross-section for $pp \to pp\eta^{\prime}$ reaction using the
parametrisation of F{\"a}ldt and Wilkin~\cite{wilkin-plb382,wilkin-prc56} fitted to the
experimental data~\cite{moskal-ijmpa22}.
For the $\eta^{\prime}$ meson the upper limit of the ratio for the excess energy range $[0,8]$~MeV
is nearly equal to values of the ratio obtained for the $\eta$ meson, whereas for larger excess
energy ranges $[8,16]$~MeV and $[16,24]$~MeV the upper limits of the ratio are lower by about
one standard deviation each.
The value of $R_{\eta}$ is $\approx~$~6.5 at excess energies larger than $\sim 16$~MeV~\cite{calen-prc58}
suggests the dominance of isovector meson exchange in the production mechanism. The  decrease of $R_{\eta}$
close to the threshold~\cite{moskal-prc79} may be explained by the different energy dependence of the
proton--proton and proton--neutron final state interactions~\cite{wilkin-priv}.
A smaller $R_{\eta^{\prime}}$ than $R_\eta$
is consistent with a possible greater role for singlet currents in $\eta^{\prime}$ production than
$\eta$ production. If there are important new dynamics in the $\eta^{\prime}$ production process relative
to eta production, a key issue is the relative phase~\cite{Fald_Wilkin_Phys_Scripta} of possible additional
exchanges compared to the isovector currents which dominate the $\eta$ production. The observed limit thus
constrains modelling of the production processes. To confirm these interesting observations and to go further,
new experimental investigations with improved statistics are required.
%pawel%%%%%%%%%%%%%%%%%%%%%%%%%%%%%%%%%%%%%%%%%%%%%%%%%%%%%%%%%%%%%%%%%%%%%%%%%%%%%%%%%%%%%%%%%%%
\section{Interactions of the \ep\ meson}
In principle, studies of $pp\to pp\, meson$ reactions permit
information about the proton--meson interaction to be gained
not only from the shape of the excitation function but also from
differential distributions of proton--proton and proton--meson invariant masses.
Therefore, in order to investigate the proton--$\eta$ interaction the COSY--11
Collaboration performed a measurement~\cite{prc69} of the proton--$\eta$
and proton--proton invariant mass distributions close to
the threshold at Q = 15.5 MeV, where the outgoing particles possess small relative velocities.
Indeed a large enhancement in the region of small proton--$\eta$ and large proton--proton relative momenta
was observed\footnote{The same enhancement was also seen in independent measurements
by the COSY--TOF Group~\cite{tof41}.}.
However, the observed effect cannot be univocally assigned
to the influence of the proton--$\eta$ interaction in the final state~\cite{fix,fix2},
since it can also  be explained by the admixture of higher partial waves
in the proton--proton system \cite{kanzo}, or by the
energy dependence of the production amplitude~\cite{deloff,ceci}.

\begin{figure}[h]
\begin{center}
 \includegraphics[height=.32\textheight]{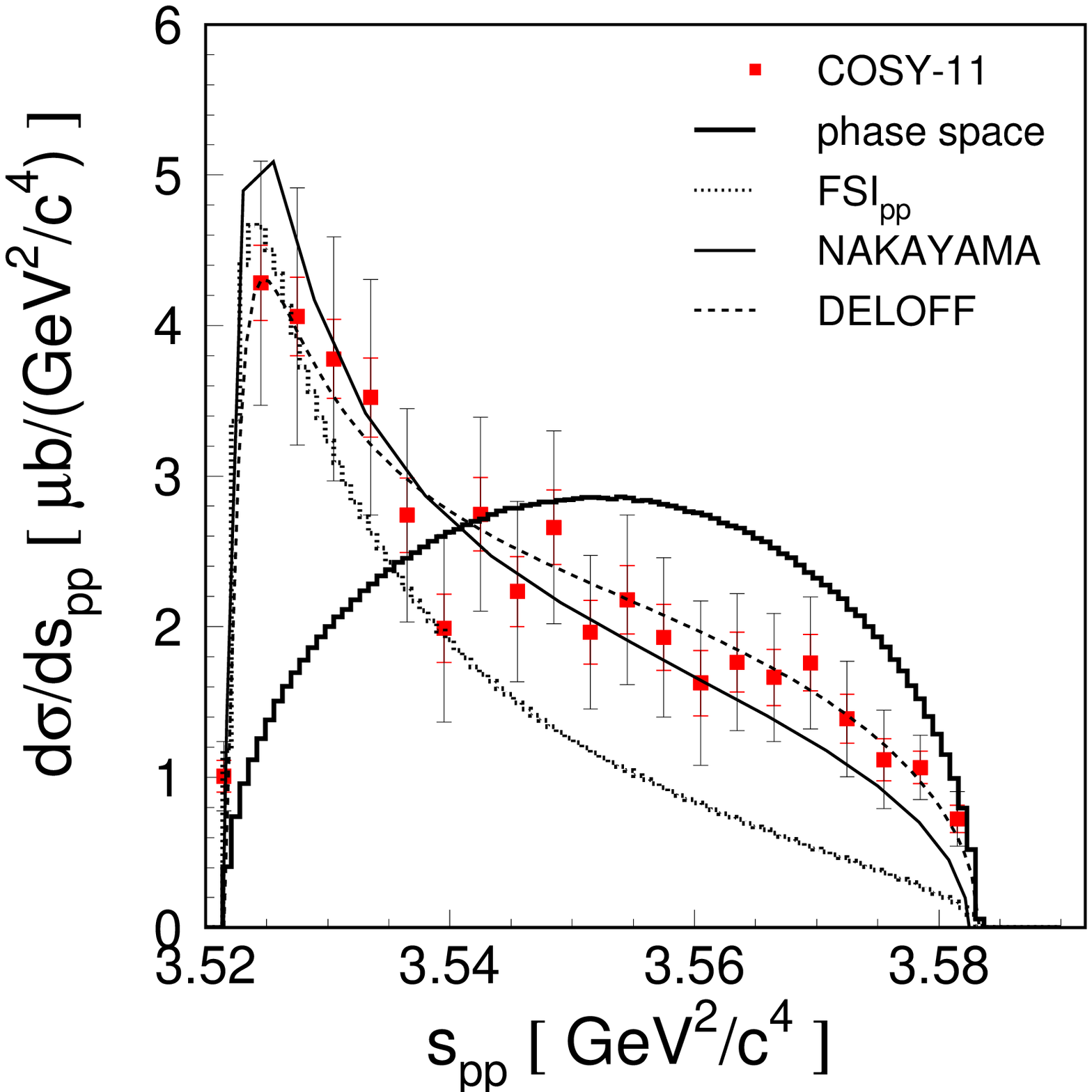}
  \includegraphics[height=.32\textheight]{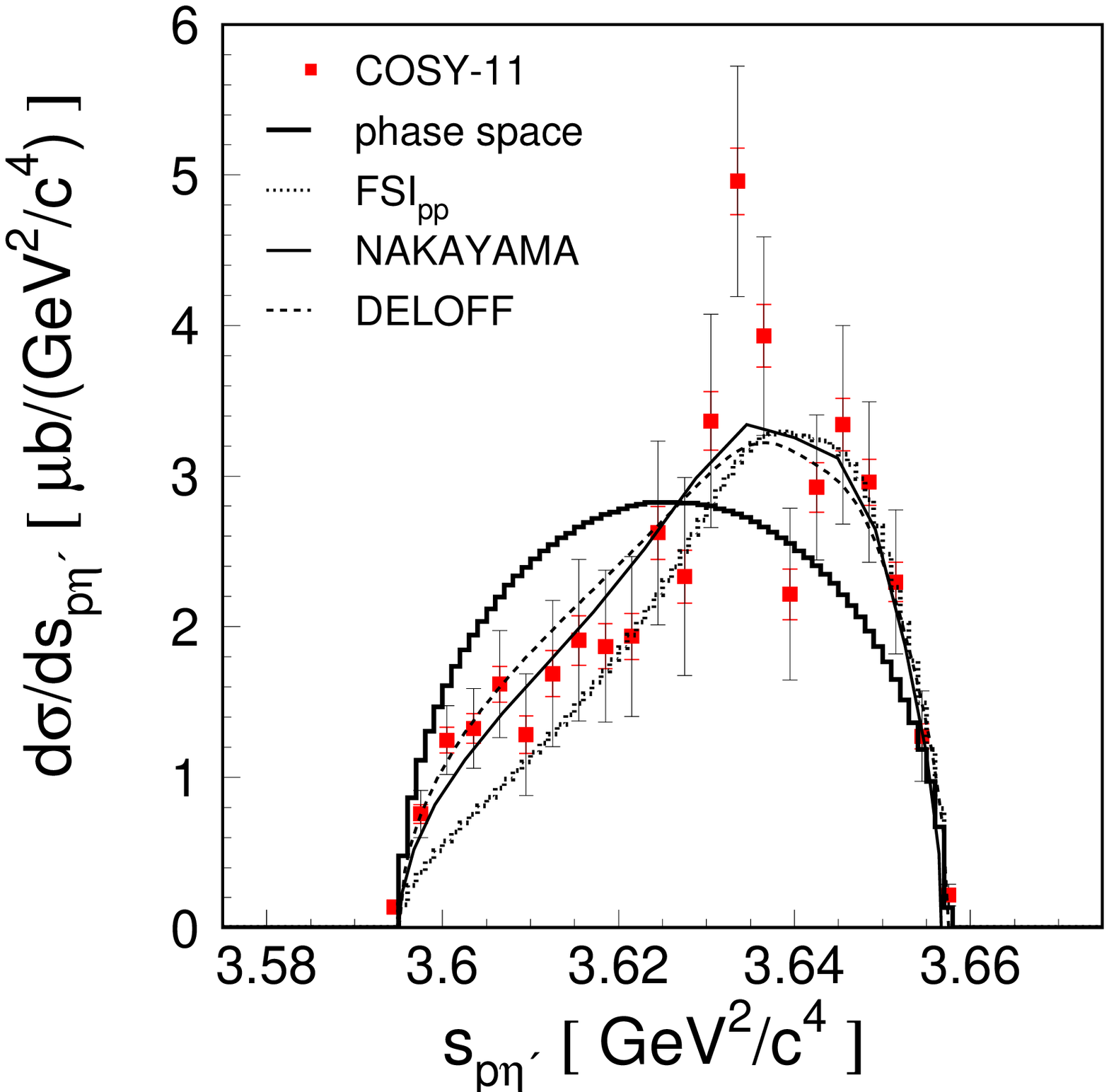}
\end{center}
        \caption{Distributions of the squared proton--proton ($s_{pp}$)
        and proton--$\eta^{\prime}$ ($s_{p\eta^{\prime}}$) invariant masses,
        for the $pp \to pp\eta^{\prime}$ reaction at the excess energy of Q = 16.4 MeV.
        The experimental data (closed squares) are
        compared to the expectation under the assumption of a homogeneously populated phase space
        (thick solid lines) and the integrals of the phase space weighted
        by the proton--proton scattering amplitude - FSI$_{pp}$ (dotted histograms).
        The solid and dashed lines correspond to calculations when taking into account contributions from higher partial waves
        and allowing for a linear energy dependence of the $^{3}P_{0} \to ^{1}\!\!S_{0}s$ partial wave amplitude, respectively.
         }
        \label{fig:norm_pp_peta}
\end{figure}

The endeavor to explain the origin of the observed enhancement
motivated the measurement of the proton--proton and proton--$\eta^{\prime}$
invariant mass distributions for the $pp \to pp\eta^{\prime}$ reaction
presented in this article.
Detailed description of measurement and data analysis resulted in invariant mass distributions for the $pp \to pp\eta^{\prime}$ reaction
is presented in \cite{pk_phd,plb_klajus}.
The absolute values of the cross-section
for the $pp \to pp\eta^{\prime}$ reaction determined as a function of $s_{pp}$ and $s_{p\eta^{\prime}}$
are shown in Fig.~\ref{fig:norm_pp_peta}.

Within the statistical and systematic error
bars both model of  Deloff~\cite{deloff} and of Nakayama et~al.\cite{kanzo}
describe the data well although they differ slightly in the predicted shapes.
This indicates that perhaps, not only higher partial waves but also the energy dependence of the production amplitude
should be taken into account.
Also, the inclusion of the proton--proton final state interaction
is not sufficient to explain the enhancement seen in the range of large $s_{pp}$ values.

Within the achieved uncertainties, the shape of the proton--proton and proton--meson
invariant mass distributions determined for the $\eta^{\prime}$ meson is essentially the same to that
established previously for the $\eta$ meson.
Since the enhancement is similar in both cases, and the strength of proton--$\eta$ and proton--$\eta^{\prime}$ interaction
is different \cite{swave, pk_phd}, one can conclude that the observed enhancement
is not caused by a proton--meson interaction.
Therefore, on the basis of the presented invariant mass distributions, it is not possible
to disentangle univocally which of the discussed models is more appropriate.
As pointed out in \cite{kanzo}, future measurements of the spin correlation coefficients
should help disentangle these two model results in a model independent way.
%eryk%%%%%%%%%%%%%%%%%%%%%%%%%%%%%%%%%%%%%%%%%%%%%%%%%%%%%%%%%%%%%%%%%%%%%%%%%%%%%%%%%%%%%%%
\section{Total width of the \ep\ meson (\epw)}
In the latest review by the Particle Data Group (PDG)~\cite{pdg}, two values
for the total width of the \ep\ meson are given.  One of these values, (0.30~$\pm$~0.09)~MeV/c$^2$, results from the average of
two measurements~\cite{Wurzinger,Binnie}, though
only in one of these experiments was $\Gamma_{\eta'}$ extracted directly
based on the mass distribution~\cite{Binnie}.
The second value (0.205~$\pm$~0.015)~MeV/c$^2$, recommended by the PDG,
is determined by fit
to altogether 51 measurements of
partial widths, branching ratios, and
combinations of particle widths obtained from integrated cross-sections~\cite{pdg}.
The result of the fit is strongly correlated with the value of
the partial width $\Gamma(\eta'\to\gamma\gamma)$,
which causes serious difficulties when the total and the partial width have to be used at the same time,
like \emph{e.g.} in studies of the gluonium content of the \ep\
meson~\cite{Biagio,Ambrosino2,Biagio2}.

The value of the total width of the $\eta'$ meson
was established directly from its  mass distribution
independently of other properties of this meson, like \emph{e.g.}
partial widths or production cross-sections.
The $\eta'$ meson was produced in proton--proton collisions via the $pp\to pp\eta'$ reaction and its
mass was reconstructed
based on the momentum vectors
of protons taking part in the  reaction.
The reader interested in the description
of the detectors and analysis procedures can find detailed informations
in Ref.~\cite{ErykPhD,eryk_prl}. The momentum of the COSY beam and the dedicated zero degree COSY--11 facility
enabled the measurement at an
excess energy  of only a fraction of an MeV above the kinematic threshold
for the $\eta^{\prime}$ meson production.
This was the most decisive factor in minimizing uncertainties of the
missing mass determination,
since at threshold the partial derivative of the missing mass
with respect to the outgoing proton momentum tends to zero.
In addition, close to threshold
the signal-to-background ratio increases due to the more rapid reduction of the phase space  for
multimeson production than for~the~$\eta^{\prime}$.
\begin{figure}[!h]
\begin{center}
\epsfig{file=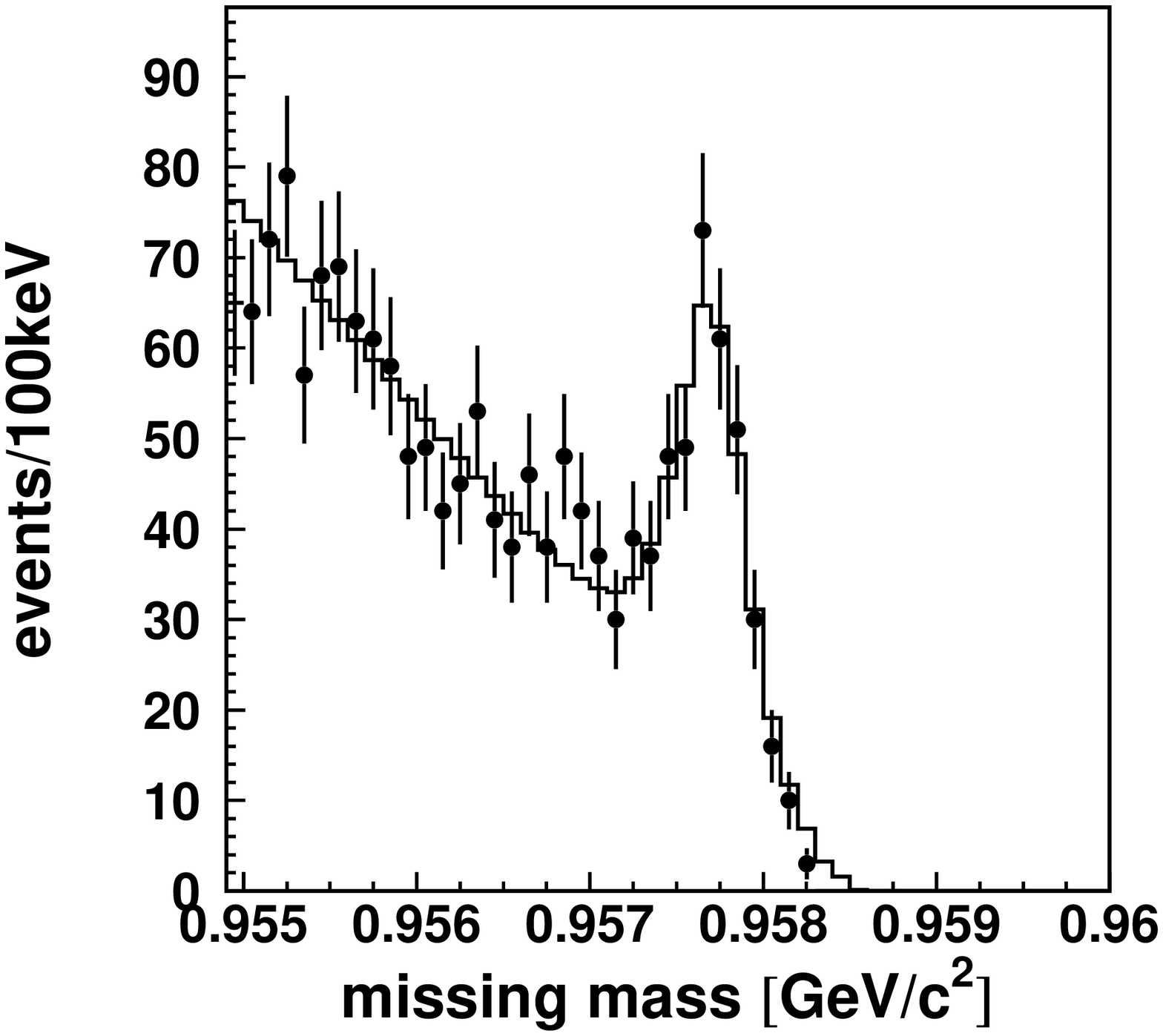,width=0.30\textwidth,angle=0}
\epsfig{file=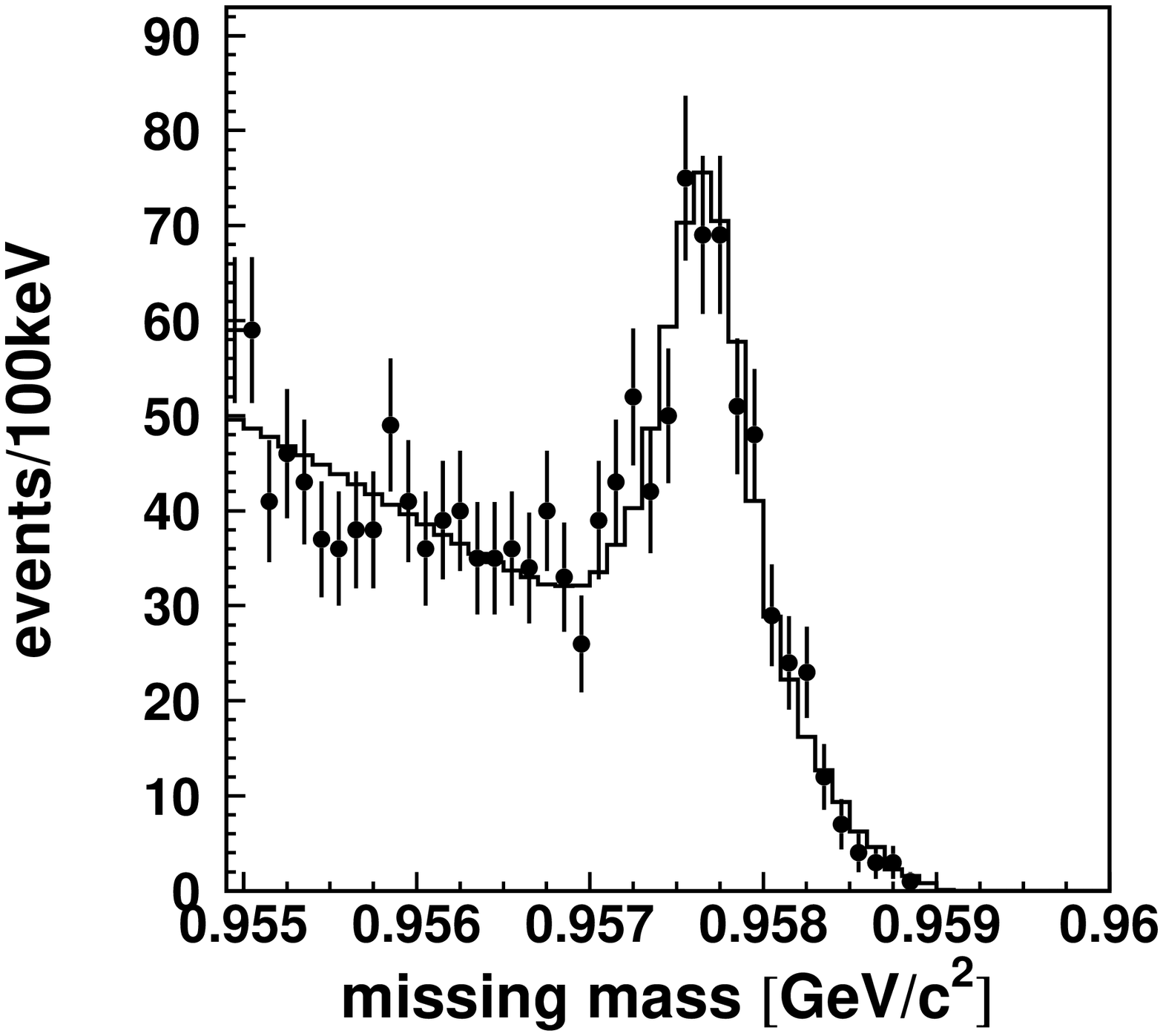,width=0.30\textwidth,angle=0}
\epsfig{file=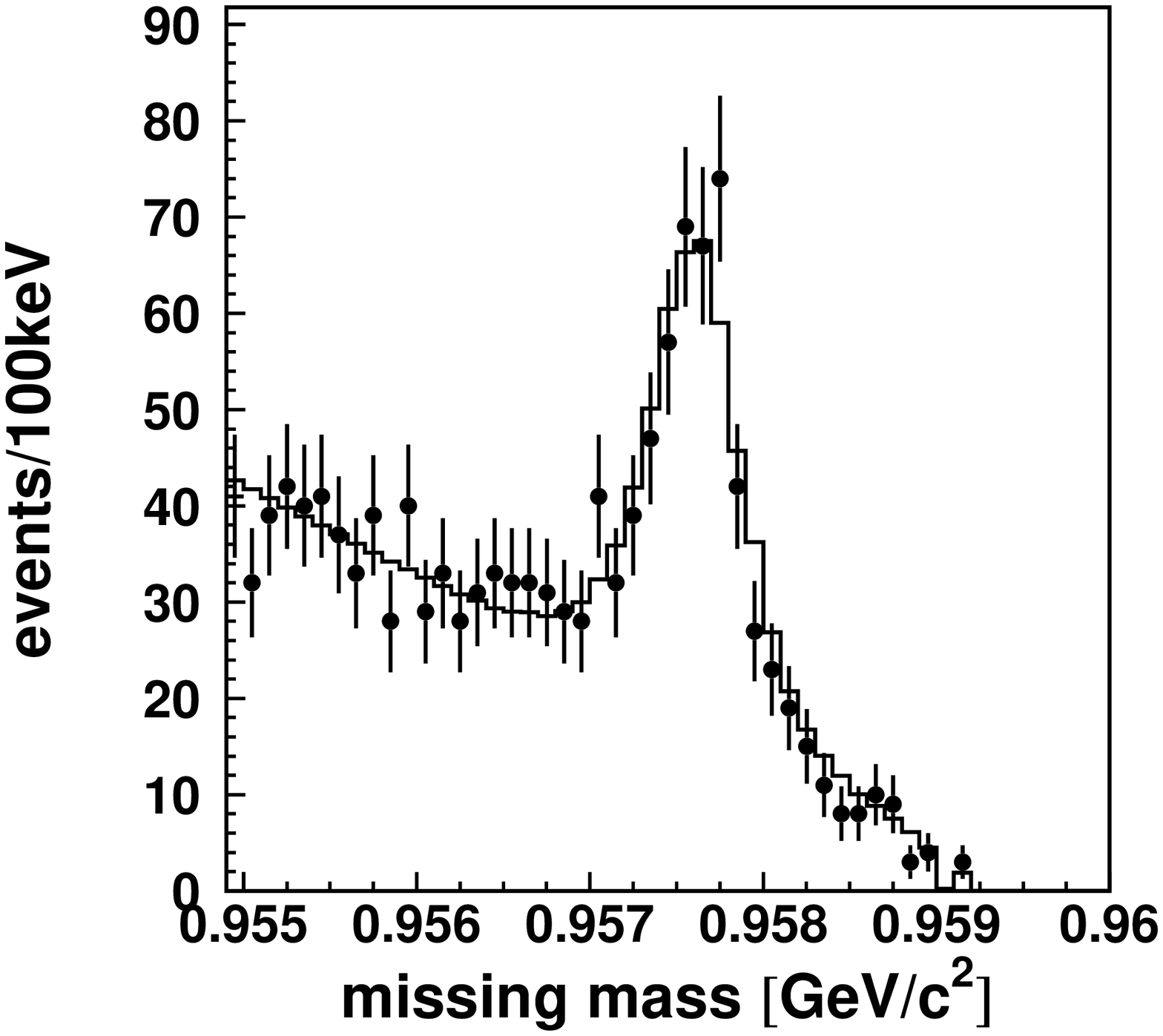,width=0.30\textwidth,angle=0}
\epsfig{file=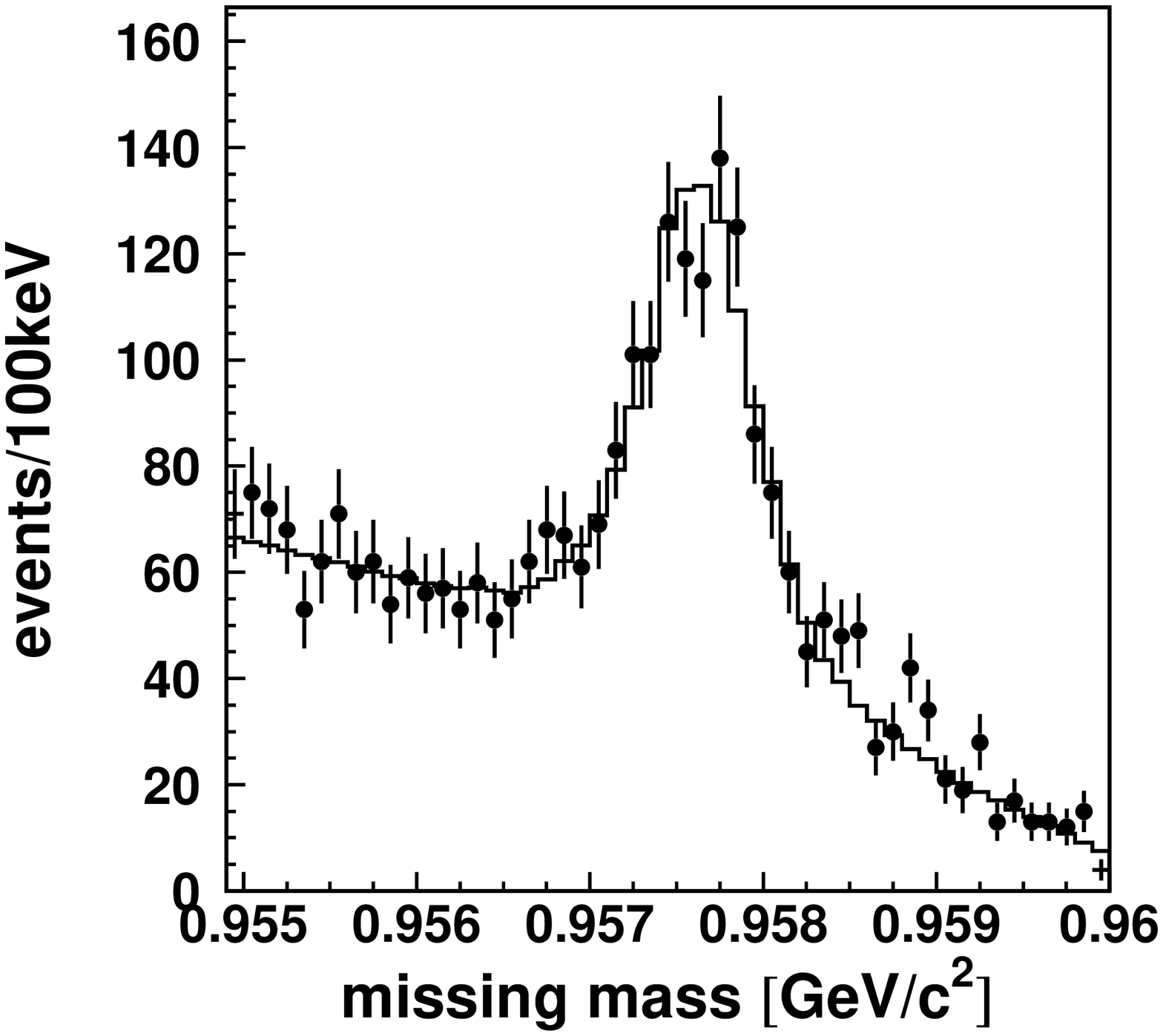,width=0.30\textwidth,angle=0}
\epsfig{file=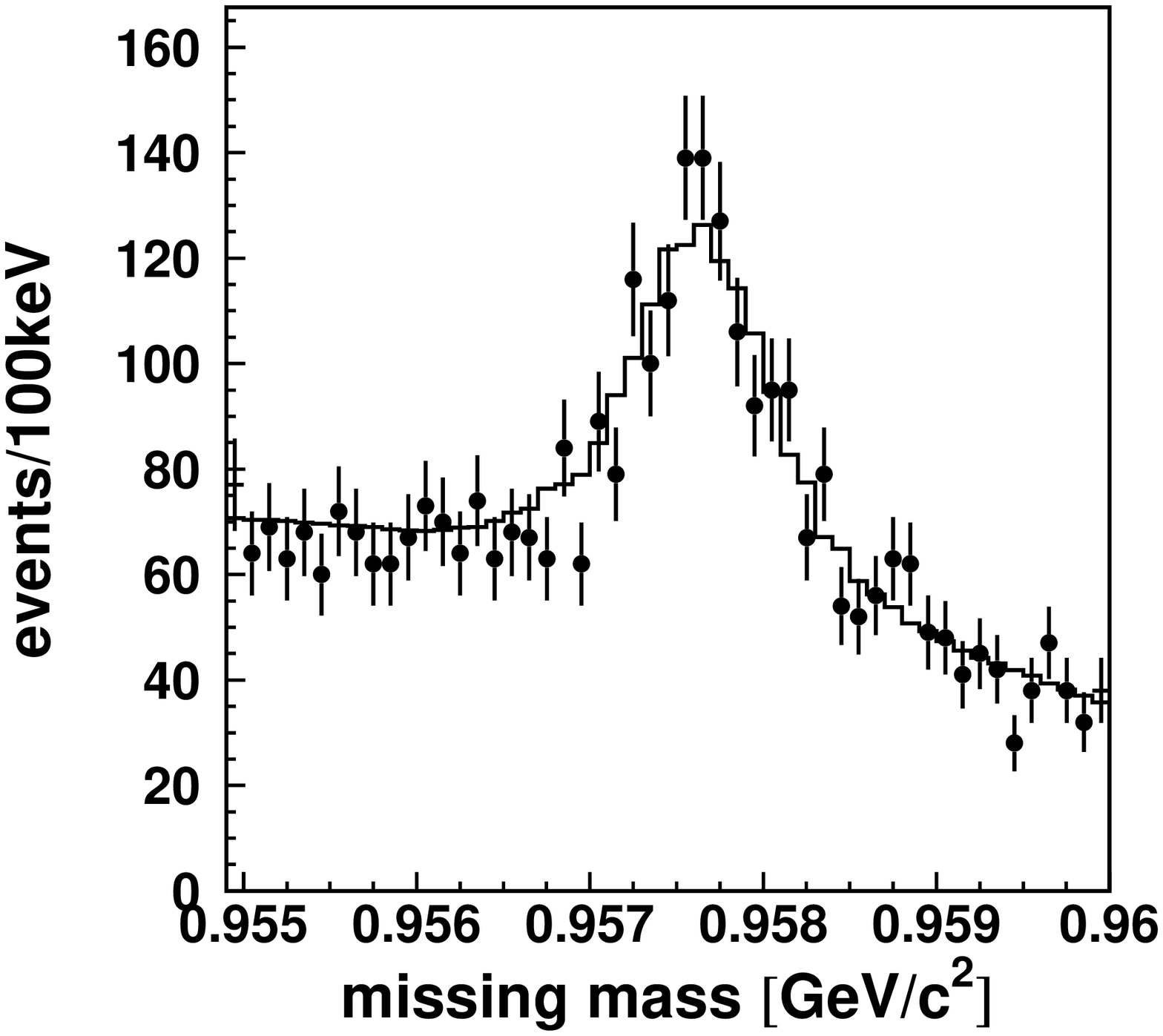,width=0.30\textwidth,angle=0}
\epsfig{file=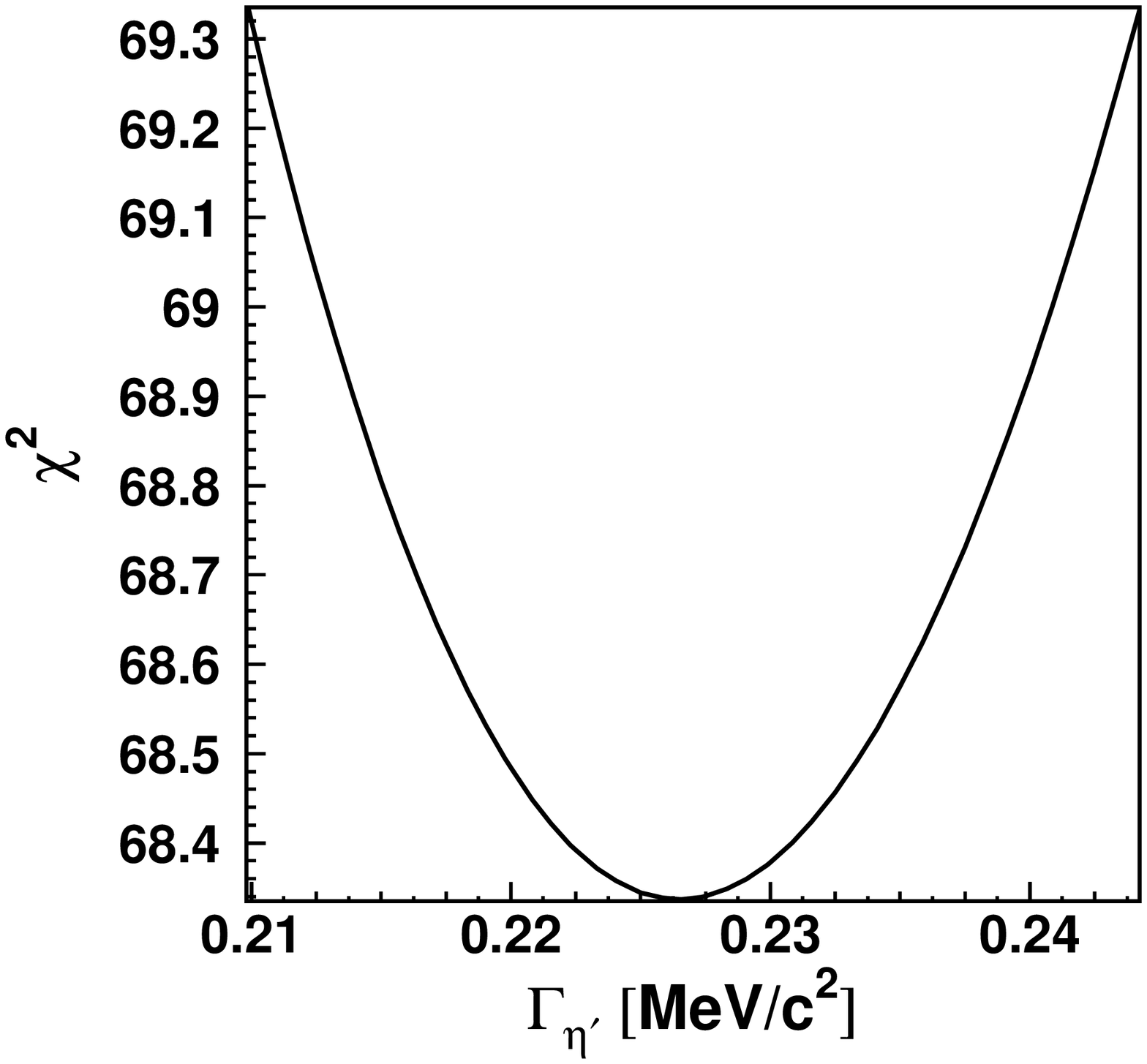,width=0.30\textwidth,angle=0}
\end{center}
\caption{
         The missing mass spectra for the $pp\to ppX$ reaction for excess energies in the CM system equal to
         0.8, 1.4, 1.7, 2.8, and 4.8~MeV
         (from left to right, top to bottom).
         The $\eta'$ meson signal is clearly visible. The experimental data are presented
         as black points, while in each plot the solid line corresponds to the sum of the Monte Carlo generated signal
         for an $\eta'$ with $\Gamma_{\eta'}=0.226$~MeV and the shifted and normalised second order polynomial obtained from
         a fit to the signal-free background region for another energy.
         The last plot (bottom right) presents  $\chi^2$ as a function of the $\Gamma_{\eta'}$.
         The minimum value corresponds to $\Gamma_{\eta'}=0.226$~MeV,
         and the range where $\chi^2=\chi^2_{min}+1$, corresponding to the
         value of the statistical error, is equal to $\pm0.017(\textrm{stat.})$.
        }
 \label{mm}
\end{figure}

The systematic error was estimated by studying the sensitivity of the result to the
variation of parameters describing the experimental conditions
in the analysis and in the simulation~\cite{ErykPhD}.
Finally, the total systematic error was estimated
as the quadratic sum of independent contributions
and is $0.014$~MeV/c$^2$.
The final missing mass spectra are presented in Fig.~\ref{mm}.
The total width of the \ep\ meson was extracted from the missing-mass spectra
and amounts to $\Gamma_{\eta'}=0.226~\pm~0.017(\textrm{stat.})$ $\pm~0.014(\textrm{syst.})$~MeV/c$^2$.
The result does not depend on knowing any of the branching ratios or partial decay widths.
The extracted $\Gamma_{\eta'}$ value is in agreement
with both previous direct determinations of this value~\cite{Binnie,Wurzinger},
and the achieved accuracy is similar to that obtained by the PDG~\cite{pdg}.
%%%%%%%%%%%%%%%%%%%%%%%%%%%%%%%%%%%%%%%%%%%%%%%%%%%%%%%%%%%%%%%%%%%%%%%%%%%%%%%%%%%%%%%%

\vspace{5mm}
The work was partially supported by the European Commission under the 7$^{\textrm{th}}$ Framework Programme
through the \emph{Research Infrastructures} action of the \emph{Capacities} Programme.
Call: FP7-INFRASTRUCTURES-2008-1, Grant Agreement No. 227431, by the PrimeNet,
by the Polish Ministry of Science and Higher Education through grants No.
1202/DFG/2007/03, 0082/B/H03/2008/34, 0084/B/H03/2008/34 and 1253/B/H03/2009/36,
by the German Research Foundation (DFG), by the FFE
grants from the Research Center J{\"u}lich, and by the virtual institute \emph{Spin and Strong
QCD} (VH-VP-231).


\begin{thebibliography}{99}
\begin{spacing}{0.5}
%introduction
\bibitem{Maier}          R.~Maier           \emph{et al., Nucl. Instrum. Methods Phys. Res.} \textbf{A390}, 1      (1997).%%CITATION = NUIMA,A390,1;%%
\bibitem{Brauksiepe}     S.~Brauksiepe      \emph{et~al., Nucl. Instrum. Methods Phys. Res.} \textbf{A376}, 397    (1996).%%CITATION = NUIMA,A376,397;%%
\bibitem{Moskal5}        P.~Moskal          \emph{et~al., Nucl. Instrum. Methods Phys. Res.} \textbf{A466}, 448    (2001).%%CITATION = NUCL-EX/0010010;%%
\bibitem{ErykPhD} E.~Czerwinski, PhD thesis, \texttt{arXiv:0909.2781 [nucl-ex]}.%%CITATION = 0909.2781;%%
\bibitem{jklaja-phd} J.~Klaja,   PhD thesis, \texttt{arXiv:0909.4399 [nucl-ex]}..%%CITATION = 0909.4399;%%
\bibitem{pk_phd} P.~Klaja,       PhD thesis, \texttt{arXiv:0907.1491 [nucl-ex]}.%%CITATION = 0907.1491;%%
%Joanna
\bibitem{cao-prc781} Xu~Cao, Xi-Guo~Lee, \emph{Phys. Rev.} {\bf C78}, 035207 (2008).%%CITATION = 0804.0656;%%
\bibitem{kampfer-ep1} L.P.~Kaptari, B.~K{\"a}mpfer, \emph{Eur. Phys. J.} {\bf A37}, 69 (2008).%%CITATION = 0804.2019;%
\bibitem{jklaja_prc} J.~Klaja {\it et al., Phys. Rev.} {\bf C81}, 035209 (2010).%%CITATION = 1003.4378;%%
\bibitem{moskal-plb474} P.~Moskal {\it et al., Phys. Lett.} {\bf B474}, 416 (2000).%%CITATION = NUCL-EX/0001001;%%
\bibitem{khoukaz-epj} A.~Khoukaz {\it et al., Eur. Phys. J.} {\bf A20}, 345 (2004).%%CITATION = NUCL-EX/0401011;%%
\bibitem{balestra-plb491} F.~Balestra {\it et al., Phys. Lett.} {\bf B491}, 29 (2000).%%CITATION = NUCL-EX/0008017;%%
\bibitem{moskal-prl80} P.~Moskal {\it et al., Phys. Rev. Lett.} {\bf 80}, 3202 (1998).%%CITATION = NUCL-EX/9803002;%%
\bibitem{hibou-plb438} F.~Hibou {\it et al., Phys. Lett.} {\bf B438}, 41 (1998).%%CITATION = NUCL-EX/9802002;%%
\bibitem{wilkin-plb382} G.~F{\"a}ldt, C.~Wilkin, \emph{Phys. Lett.} {\bf B382}, 209 (1996).%%CITATION = PHLTA,B382,209;%%
\bibitem{wilkin-prc56} G.~F{\"a}ldt, C.~Wilkin, \emph{Phys. Rev.} {\bf C56}, 2067 (1997).%%CITATION = NUCL-TH/9704056;%%
\bibitem{moskal-ijmpa22} P.~Moskal {\it et al., Int. J. Mod. Phys.} {\bf A22}, 305 (2007).%%CITATION = HEP-EX/0609035;%%
\bibitem{calen-prc58} H.~Cal{\'e}n {\it et al., Phys. Rev.} {\bf C58}, 2667 (1998).%%CITATION = PHRVA,C58,2667;%%
\bibitem{moskal-prc79} P.~Moskal {\it et al., Phys. Rev.} {\bf C79}, 015208 (2009).%%CITATION = 0807.0722;%%
\bibitem{wilkin-priv} C.~Wilkin, private communication (2008).
\bibitem{Fald_Wilkin_Phys_Scripta} G.~F{\"a}ldt, C.~Wilkin, \emph{Phys. Scr.} {\bf 64}, 427 (2001).%%CITATION = NUCL-TH/0104081;%%
%Pawel
\bibitem{prc69} P. Moskal \emph{et al., Phys. Rev.} {\bf C69}, 025203 (2004).%%CITATION = NUCL-EX/0307005;%%
\bibitem{tof41} M. Abdel-Bary \emph{et al., Eur. Phys. J.} {\bf A16}, 127 (2003).%%CITATION = NUCL-EX/0205016;%%
\bibitem{fix} A. Fix, H. Arenh{\"o}vel, \emph{Phys. Rev.} {\bf C69}, 014001 (2004).%%CITATION = NUCL-TH/0310034;%%
\bibitem{fix2} A. Fix, H. Arenh{\"o}vel, \emph{Nucl. Instrum. Methods Phys. Res.} {\bf A697}, 277 (2002).
\bibitem{kanzo} K. Nakayama \emph{et al., Phys. Rev.} {\bf C68}, 045201 (2003).%%CITATION = NUCL-TH/0302061;%%
\bibitem{deloff} A. Deloff, \emph{Phys. Rev.} {\bf C69}, 035206 (2004).%%CITATION = NUCL-TH/0309059;%%
\bibitem{ceci} S. Ceci, A. {\v{S}}varc, B. Zauner, \emph{Acta Phys. Pol. B Proc. Supp.}  {\bf 2}, 157 (2009).
\bibitem{plb_klajus} P. Klaja \emph{et al., Phys. Lett.} {\bf B684}, 11 (2010).%%CITATION = 1001.5174;%%
\bibitem{swave} P. Moskal \emph{et al., Phys. Lett.} {\bf B482}, 356 (2000).%%CITATION = NUCL-EX/0004006;%%
%Eryk
\bibitem{pdg}            C.~Amsler          \emph{et~al., Phys. Lett.}              \textbf{B667}, 1      (2008).%%CITATION = PHLTA,B667,1;%%
\bibitem{Wurzinger}      R.~Wurzinger       \emph{et~al., Phys. Lett.}             \textbf{B374}, 283    (1996).%%CITATION = PHLTA,B374,283;%%
\bibitem{Binnie}         D.M. Binnie       \emph{et~al., Phys. Lett. }            \textbf{B83},  141     (1979).%%CITATION = PHLTA,B83,141;%%
\bibitem{Biagio}         B.~Di~Micco,                      \emph{Acta. Phys. Pol. B Proc. Supp.} \textbf{2},   63     (2009).
\bibitem{Ambrosino2}     F.~Ambrosino       \emph{et~al., J. High Energy Phys.}                     \textbf{07},  105    (2009).%%CITATION = 0906.3819;%%
\bibitem{Biagio2}        B.~Di~Micco,                      \emph{Eur. Phys. J.}          \textbf{A38},  129    (2008).%%CITATION = EPHJA,A38,129;%%
\bibitem{eryk_prl}  E.~Czerwinski, P.~Moskal \emph{et al., Phys. Rev. Lett.} {\bf 105}, 122001 (2010).%%CITATION = PRLTA,105,122001;%%
 
\end{spacing}
\end{thebibliography}
\end{document}